\documentclass[aps,prl,twocolumn,superscriptaddress]{revtex4}
\usepackage{epsf,graphicx}
\usepackage{amssymb}
\usepackage{amsmath}

\usepackage{color}
\usepackage{ulem}
\begin{document}

\title{Continuous ferromagnetic quantum phase transition on an anisotropic Kondo lattice}
\author{Jialin Chen}
\affiliation{Beijing National Laboratory for Condensed Matter Physics and Institute of
Physics, Chinese Academy of Sciences, Beijing 100190, China}
\affiliation{University of Chinese Academy of Sciences, Beijing 100049, China}
\author{Jiangfan Wang}
\affiliation{Beijing National Laboratory for Condensed Matter Physics and Institute of
Physics, Chinese Academy of Sciences, Beijing 100190, China}
\author{Danqing Hu}
\affiliation{Beijing National Laboratory for Condensed Matter Physics and Institute of
	Physics, Chinese Academy of Sciences, Beijing 100190, China}
\author{Yi-feng Yang}
\email[]{yifeng@iphy.ac.cn}
\affiliation{Beijing National Laboratory for Condensed Matter Physics and Institute of
Physics, Chinese Academy of Sciences, Beijing 100190, China}
\affiliation{University of Chinese Academy of Sciences, Beijing 100049, China}
\affiliation{Songshan Lake Materials Laboratory, Dongguan, Guangdong 523808, China}
\date{\today}

\begin{abstract}
Motivated by the recent discovery of ferromagnetic quantum criticality in the heavy fermion compound CeRh$_6$Ge$_4$, we develop a numerical algorithm of infinite projected entangled pair states for the anisotropic ferromagnetic Kondo-Heisenberg model in two dimensions and study the ferromagnetic quantum phase transitions with varying magnetic and hopping  anisotropy. Our calculations reveal a continuous ferromagnetic quantum phase transition in the large anisotropic region and first-order quantum phase transitions for smaller anisotropy. Our results  highlight the importance of magnetic anisotropy on ferromagnetic quantum criticality in Kondo lattice systems  and provide a possible explanation for the experimental observation in CeRh$_6$Ge$_4$ with a quasi-one-dimensional magnetic structure. Our work opens the avenue for future studies of the rich Kondo lattice physics using state-of-the-art tensor network approaches.
\end{abstract}

\maketitle

A quantum phase transition (QPT) refers to a phase transition at zero temperature. A quantum critical point (QCP) marks a continuous QPT and often induces exotic scaling behaviors in physical properties at finite temperatures \cite{Sachdev1999Quantum,Yang2020}. The recent discovery of a ferromagnetic (FM) QCP in the clean heavy fermion compound CeRh$_6$Ge$_4$ under hydrostatic pressure \cite{Shen2020Strange} has stimulated intensive interest in clarifying its underlying mechanism. According to prevailing theories, FM quantum phase transitions in heavy fermion systems, as manifested in UGe$_2$ \cite{Taufour2010Tricritical} and URhAl \cite{Shimizu2015Unusual}, are typically first order because of some soft modes associated with the electron Fermi surfaces \cite{Kirkpatrick2017The, Brando2016Metallic}. A continuous FM QPT had previously been observed in YbNi$_4$(P$_{0.92}$As$_{0.08}$)$_2$ \cite{Krellner2011Ferromagnetic, Alexander2013Ferromagnetic}, but might be due to disorder introduced by chemical substitution. However, this cannot explain the observation in the stoichiometric CeRh$_6$Ge$_4$, which demands a novel theoretical understanding. 

Very lately, it was suggested that the noncentrosymmetric crystal structure and strong spin-orbit coupling (SOC) may induce a gap in the soft modes and invalidate the argument for a first-order FM QPT \cite{Kirkpatrick2020Ferromagnetic}. But this claim was soon disputed by a fully self-consistent non-perturbative analysis finding that backscattering processes ignored in previous work can always contribute a non-analytic term in the free energy and cause a first-order transition for arbitrarily complex SOC \cite{miserev2022instability}. Alternatively, it was also pointed out that magnetic anisotropy might be important \cite{Shen2020Strange, Komijani2018Model}, since both CeRh$_6$Ge$_4$ \cite{Shen2020Strange} and YiNi$_4$(P$_{0.92}$As$_{0.08}$)$_2$ \cite{Krellner2011Ferromagnetic, Alexander2013Ferromagnetic} have a quasi-one-dimensional magnetic structure. The reduced dimensionality might enhance magnetic fluctuations and tune the first-order QPT to a continuous one \cite{Kitagawa2013Ferromagnetic}. Indeed, latest calculations based on large-$N$ Schwinger boson approximation confirmed this expectation and yielded a unified phase diagram showing the variation from first to second-order QPTs  with increasing magnetic anisotropy \cite{wang2021unified}. But all these results require further scrutiny by more accurate numerical approaches.

Tensor network variational methods have been developed in recent years for complex quantum many-body systems and do not suffer from the negative sign problem in quantum Monte Carlo simulations \cite{Orus2019Tensor}. For one-dimensional (1D) or stripe lattices \cite{Stoudenmire2012Studying,Xie2015}, the density matrix renormalization group (DMRG) based on matrix product states \cite{White1992Density} provides a useful tool of reference. Its extension to 2D has led to the finite-size projected entangled pair states (PEPS) \cite{verstraete2004renormalization} and its infinite version iPEPS \cite{Jordan2008Classical}, which were initially designed for spin and boson systems but also have a potential for fermionic systems \cite{Wolf2006Violation, Gioev2006Entanglement, Miao2021Eigenstate}. Their preliminary applications include the study of stripe orders and its competing relation to the uniform $d$-wave state in the doped 2D Hubbard model \cite{Zheng2017Stripe, Ponsioen2019Period} and $t$-$J$ model \cite{Corboz2011Stripes, Corboz2014Competing}, calculations of Fermi surfaces \cite{mortier2021tensor}, finite-temperature state \cite{Czarnik2016Variational} and excitation \cite{ponsioen2021automatic, Schneider2021Simulating} of the 2D Hubbard model, and determination of the ground state of two-band or three-flavor Hubbard models \cite{Benedikt2021A}. But so far, these methods have not been applied to the Kondo lattice systems consisting of both fermionic and spin degrees of freedom.

In this work, we extend iPEPS to the ferromagnetic Kondo lattice in 2D with anisotropic magnetic interactions and hopping parameters, and explore the FM ground state and its suppression with increasing Kondo coupling. This allows us to provide confirmative numerical evidence for the property of the FM QPTs and establish a global phase diagram tuned by the magnetic and hopping anisotropy. We find the transition to be continuous for large anisotropy close to the quasi-1D limit and first-order for less anisotropy. Although our model is not realistic for a particular material, our findings still provide a possible generic mechanism for understanding first- or second-order FM QPTs in clean heavy fermion ferromagnets. The resulting phase diagram agrees well with previous theoretical prediction based on the large-$N$ Schwinger boson approximation \cite{wang2021unified} and highlights the importance of quasi-1D magnetic structure on the FM QCP. Our work opens an opportunity for future exploration of the rich physics in Kondo lattice systems using state-of-the-art tensor network approaches.

\begin{figure}[t]
\begin{center}
\includegraphics[width=8.6cm]{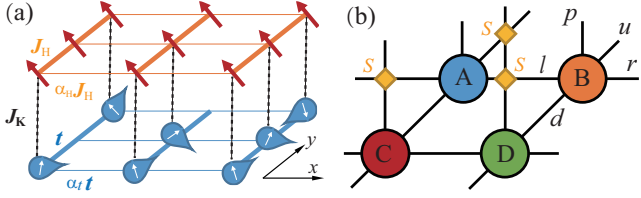}
\end{center}
\caption{(a) Illustration of the anisotropic FM Kondo-Heisenberg model studied in this work. The orange arrows represent local spins and the blue droplets with white arrows denote conduction electrons. The reduced width of the solid lines along the $x$-direction reflects the reduced strength of the Heisenberg exchange interaction and the hopping parameter due to lattice anisotropy. (b) Illustration of the $2\times 2$ unit cell of the iPEPS ansatz with four bulk tensors: A, B, C, D. Each has one physical bond ($p$) and four virtual bonds ($l$, $u$, $r$, $d$)  as noted for B. The yellow diamonds represent the 4-leg swap gates $S$ used for the fermionization of iPEPS, which is introduced at the crossing point of two legs \cite{Corboz2010Simulation, Corboz2009Fermionic}.}
\label{fig1}
\end{figure}

We start with the following 2D ferromagnetic Kondo-Heisenberg model with anisotropic exchange interactions and hopping parameters:
\begin{align}{\label{Hamiltonian}}
H&=-t\sum_{\mathbf{r}\sigma} \left(\alpha_t c^\dagger_{\mathbf{r}\sigma} c_{\mathbf{r}+\mathbf{x}_0,\sigma} +  c^\dagger_{\mathbf{r}\sigma} c_{\mathbf{r}+\mathbf{y}_0,\sigma} + \text{H.c.} \right) \\
  &+ J_\text{K} \sum_\mathbf{r} \mathbf{s}_\mathbf{r} \cdot \mathbf{S}_\mathbf{r}  \nonumber 
  +J_\mathrm{H} \sum_\mathbf{r} \left( \alpha_\mathrm{H} \mathbf{S}_{\mathbf{r}}\cdot \mathbf{S}_{\mathbf{r}+\mathbf{x}_0} + \mathbf{S}_{\mathbf{r}} \cdot \mathbf{S}_{\mathbf{r}+\mathbf{y}_0} \right), 
\end{align}
where $c^\dagger_{\mathbf{r}\sigma}$ creates an electron of spin $\sigma$ at position $\mathbf{r}=(x,y)$, $\mathbf{x}_0=(1,0)$ and $\mathbf{y}_0=(0,1)$ denote two basis vectors of the 2D lattice, $t$ is the hopping amplitude along $y$-axis, $J_\mathrm{K}$ is the Kondo coupling between local spin $\mathbf{S}_\mathbf{r}$ and conduction electron spin $\mathbf{s}_\mathbf{r}=\sum_{\alpha\beta}c_{\mathbf{r}\alpha}^{\dag}\frac{\vec{\sigma}_{\alpha\beta}} {2}c_{\mathbf{r}\beta}$,  $J_\mathrm{H}<0$ denotes the ferromagnetic Heisenberg exchange interaction between neighboring local spins within the Kondo chains along the $y$-axis, and $0\le \alpha_{t/\mathrm{H}}\le 1$ reflect the hopping or magnetic anisotropy along $x$-axis due to the anisotropic lattice structure. An illustration of the model is plotted in Fig.~\ref{fig1}(a). For simplicity, $t$ is set to unity as the energy unit. The conduction electron occupation is one per site for the current model in order to fully screen the local spins and obtain the FM QPT at large $J_\mathrm{K}$. The ferromagnetic Heisenberg $J_\mathrm{H}$ is included explicitly in order to overcome the induced antiferromagnetic (AFM) RKKY interaction.

\begin{figure}[t]
	\begin{center}
		\includegraphics[width=8.6cm]{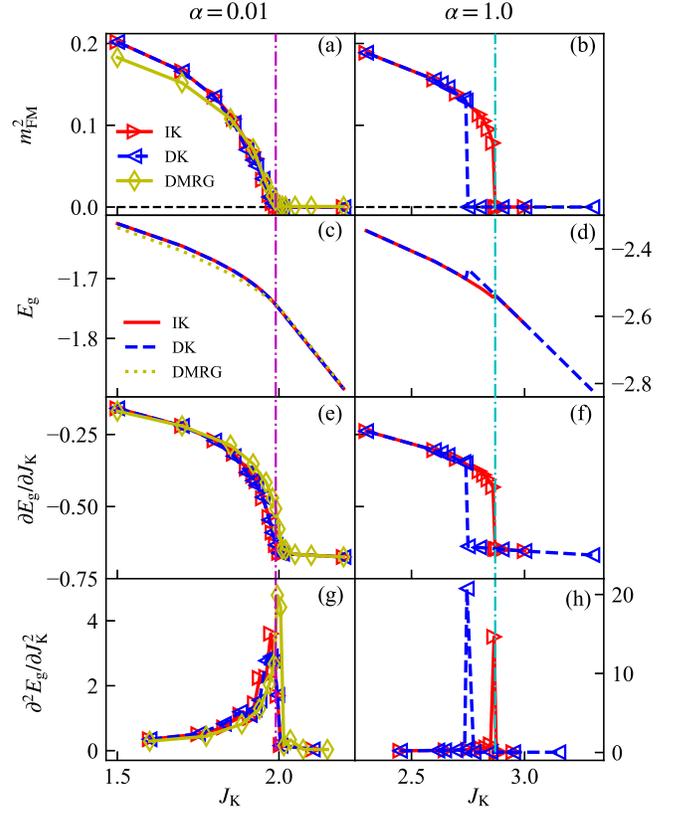}
	\end{center}
	\caption{Comparison of the FM QPTs in the isotropic ($\alpha=1$) and quasi-1D ($\alpha=0.01$) limits based on the $J_\mathrm{K}$ dependence of (a)(b) $m^2_\mathrm{FM}$, the FM order parameter  defined in Eq.~(\ref{correlation_m}), (c)(d) $E_\mathrm{g}$, the ground state energy per site, (e)(f) the derivative $\partial E_\mathrm{g} / \partial J_\mathrm{K} = \langle \mathbf{s}_i \cdot \mathbf{S}_i \rangle$, and (g)(h) the second derivative $\partial^2 E_\mathrm{g} / \partial J_\mathrm{K}^2$. The red solid and blue dashed lines represent two different routes of calculations in the hysteresis strategy with gradually increasing (IK) and decreasing (DK) $J_\mathrm{K}$, respectively. The yellow solid or dotted lines are DMRG results on the 1D Kondo chain for comparison. The purple and cyan dash-dotted vertical lines mark the continuous or first-order (energy crossing) transition point, respectively. The virtual bond dimension of iPEPS is set to $D=12$.}
	\label{fig2}
\end{figure}

We solve the above model using the fermionic iPEPS \cite{Corboz2010Simulation, Benedikt2021A} consisting of 2D unit cells with rank-5 bulk tensors arranged periodically. To conform with geometric anisotropy and possible AFM correlation due to the induced RKKY interaction, we have chosen a $2\times2$ unit cell with 4 different bulk tensors as shown in Fig.~\ref{fig1}(b). Each bulk tensor has a physical bond $p$ of dimension $d_p=8$ representing the dimension of the local Hilbert space formed by the direct product of 4 electron configurations and 2 local spin configurations. The dimension $D$ of its four virtual bonds ($l$, $u$, $r$, $d$) controls the upper limit of the entanglement among neighboring sites. Each basis of the tensor is given a parity index $p=(-1)^{n_f}$, where $n_f$ is the fermionic particle number in that basis, so that the anticommutation relation of electrons can be represented using swap gates and parity-invariant tensors with the same leading cost as bosonic iPEPS \cite{Corboz2010Simulation, Corboz2009Fermionic}. The algorithm is then optimized using simple update \cite{Corboz2010Simulation, Czarnik2016Variational} with a newly developed energy estimator \cite{Schneider2021Simulating} and random initial or saved iPEPS. The calculations converge as the energy achieves the accuracy of $\Delta E=10^{-12}$. We have also applied the fast full update algorithm \cite{Phien2015Infinite} and the results are qualitatively the same \cite{SM}. The expectation values of physical properties are evaluated using the corner transfer matrix renormalization group (CTMRG) \cite{Orus2009Simulation, Corboz2014Competing} with the environment bond dimension $\chi=D^2$. The results can in principle be improved with increasing $D$, but the computational cost explodes as $D^{12}$. Fortunately, we find $D=12$, which is the limit of our numerical capability, can already give some convergent results for the present model. 

To focus on the overall variation of the FM QPTs, we only show the data for $J_\mathrm{H}=-1.0$ and fix $\alpha_t=\alpha_\mathrm{H}=\alpha$. The results for other parameter regions  will only be discussed briefly. The ground state of the model is FM for small $J_\mathrm{K}$ and paramagnetic (PM) for sufficiently large $J_\mathrm{K}$ due to Kondo screening. The magnetization in the FM phase can be characterized by the spin correlation function along the $y$-axis (or $x$-axis) in the thermodynamic limit \cite{Hasik2019Optimization},
\begin{equation}
m^2_\mathrm{FM}=\lim_{L\to\infty}\frac{1}{L} \sum^{L}_{y=1}\left\langle \mathbf{S}_{\left( 0,0\right)} \cdot \mathbf{S}_{\left( 0,y\right)} \right\rangle.
 \label{correlation_m}
\end{equation}
We first obtain the two limits by optimizing some random initial states at sufficiently small or large $J_\mathrm{K}$. The ground states in between are evaluated iteratively by gradually increasing or decreasing $J_\mathrm{K}$, respectively. The nature of the FM QPT is then determined using a hysteresis calculation strategy \cite{Orus2009First} by comparing the two results. For $\alpha=0$, the model is an array of decoupled Kondo chains, which we have also calculated using DMRG in an ITensor implementation \cite{itensor} with the length $L=100$, the virtual bond dimension $\chi'=1800$, and the truncation error $10^{-6}$. 

Figure \ref{fig2} compares the iPEPS results for two extreme cases, namely, the isotropic limit ($\alpha=1.0)$ and the quasi-1D limit ($\alpha=0.01$). The red and blue symbols represent the results from increasing and decreasing $J_\mathrm{K}$, respectively. The quasi-1D results are compared with DMRG calculations on the Kondo chain and find reasonable consistency. A hysteresis is clearly seen in the order parameter $m^2_\mathrm{FM}$ for $\alpha=1.0$ but missing for $\alpha=0.01$, suggesting a first-order transition in the isotropic model and a continuous one in the quasi-1D limit. This is further supported by the evolution of the ground state energy $E_\mathrm{g}$, which overlaps between increasing and decreasing $J_\mathrm{K}$ for $\alpha=0.01$ but shows a clear difference in the hysteresis region for $\alpha=1.0$. The energy curves for $\alpha=0.01$ are also consistent with the DMRG one. For $\alpha=1.0$, the true transition point is typically assigned to the crossing point of two energy curves, which, quite interestingly, coincides with the right boundary of the QPT here.

Theoretically, the order of a QPT can be classified by the discontinuity in the derivatives of the free energy or, at zero temperature, the ground state energy, with respect to the tuning parameter. For the Kondo lattice model, one has $\partial E_\mathrm{g} / \partial J_\mathrm{K}=\langle \mathbf{s}_i \cdot \mathbf{S}_i \rangle $, which represents the Kondo screening between local spins and conduction electrons \cite{Raczkowski2020Phase}. Thus, the order of the QPT also reflects a jump or continuous change of the Kondo screening across the transition. In Figs.~\ref{fig2}(e) and \ref{fig2}(f), we see in the first derivative a similar hysteresis and jump for $\alpha=1$ and continuous change for $\alpha=0.01$, confirming that the FM QPT is first order for the isotropic model and continuous in the quasi-1D case. For $\alpha=0.01$, Fig. \ref{fig2}(g) further reveals a discontinuity (divergence) in the second derivative $\partial^2 E_\mathrm{g} /\partial J^2_\mathrm{K}$ at the QCP, implying that the transition is second order.

\begin{figure}[t]
	\begin{center}
		\includegraphics[width=8.0cm]{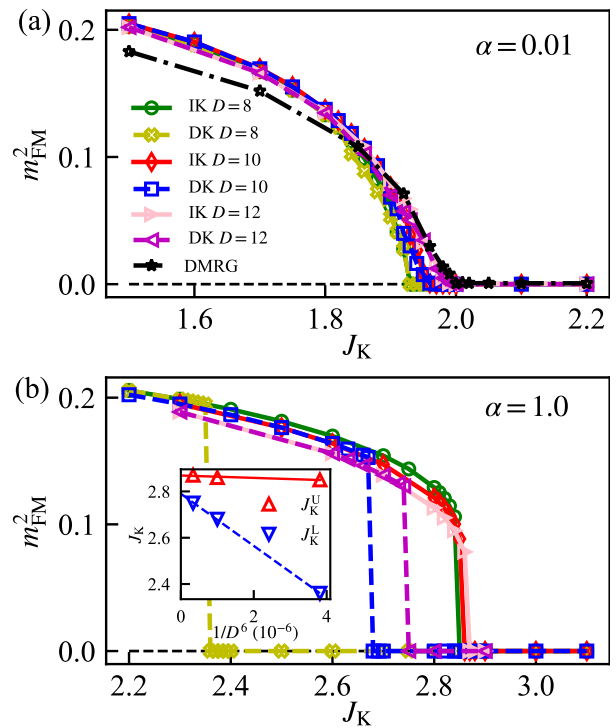}
	\end{center}
	\caption{Evolution of $m^2_\mathrm{FM}$-$J_\mathrm{K}$ curves for (a) $\alpha=0.01$ and (b) $\alpha=1.0$ with varying iPEPS virtual bond dimensions $D=8$, $10$, $12$. The DMRG results (black dash-dotted line) on the Kondo chain are also shown in (a) for comparison. The inset of (b) plots the extrapolated lower (upper) hysteresis boundary $J^\mathrm{L}_\mathrm{K}$ ($J^\mathrm{U}_\mathrm{K}$) as $D^{-6}$ for $\alpha=1.0$.}
	\label{fig3}
\end{figure}

One may have noticed that DMRG calculations for the Kondo chain ($\alpha=0$) yield a slightly larger critical $J_\mathrm{K}$ than the iPEPS for $\alpha=0.01$. Apart from the small variation in $\alpha$, we also attribute such a difference to the error arising from the finite virtual bond dimension $D=12$ used in our iPEPS calculations. This, unfortunately, already reaches the limit of our computational capability for the Kondo-Heisenberg model with a local physical dimension $d_p=8$. To have a better idea of the overall tendency of the critical $J_\mathrm{K}$ with varying bond dimension $D$, we compare in Fig.~\ref{fig3}(a) the $m^2_\mathrm{FM}$ curves for $D=8$, 10, 12. The results are qualitatively the same, with only the QCP moving slightly towards higher values of $J_\mathrm{K}$ and approaching the DMRG prediction with increasing $D$. We thus conclude that, to the best of our numerical capability, the iPEPS results are consistent with DMRG and the FM QPT is continuous in the quasi-1D FM Kondo lattice. For comparison, Fig.~\ref{fig3}(b) also shows the results of $D=8$, 10, 12 for $\alpha=1.0$. Interestingly, while the upper boundary remains almost unchanged, the lower phase boundary moves rapidly towards a larger $J_\mathrm{K}$ with increasing $D$. Their overall variations are summarized in the inset and plotted against $D^{-6}$. The difference between the two boundaries is significantly reduced for larger $D$, indicating a much narrower hysteresis region for the exact solutions in the limit of infinite bond dimension. Nevertheless, one still finds a finite difference in their extrapolated values and the nature of the first-order FM QPT is qualitatively unchanged for $\alpha=1.0$.

\begin{figure}[t]
\begin{center}
\includegraphics[width=8.4cm]{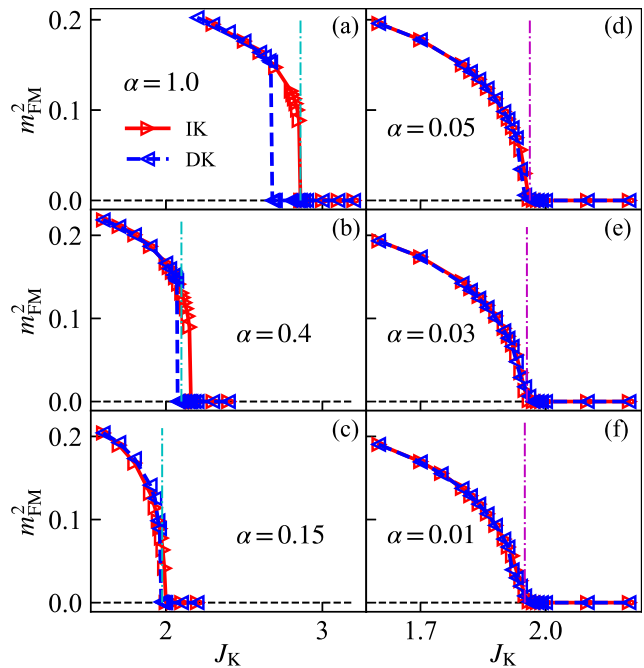}
\end{center}
\caption{Comparison of $m^2_\mathrm{FM}$-$J_\mathrm{K}$ curves for $\alpha=1.0$, 0.4, 0.15, 0.05, 0.03, 0.01. The red solid and blue dashed lines represent the results for increasing (IK) and decreasing (DK) $J_\mathrm{K}$, respectively. The virtual bond dimension is set to $D=10$ to reduce the computational cost. The dash-dotted vertical lines mark the continuous or first-order (energy crossing) transition point for each $\alpha$, respectively.}
\label{fig4}
\end{figure}

\begin{figure}[ptb]
	\begin{center}
		\includegraphics[width=8.4cm]{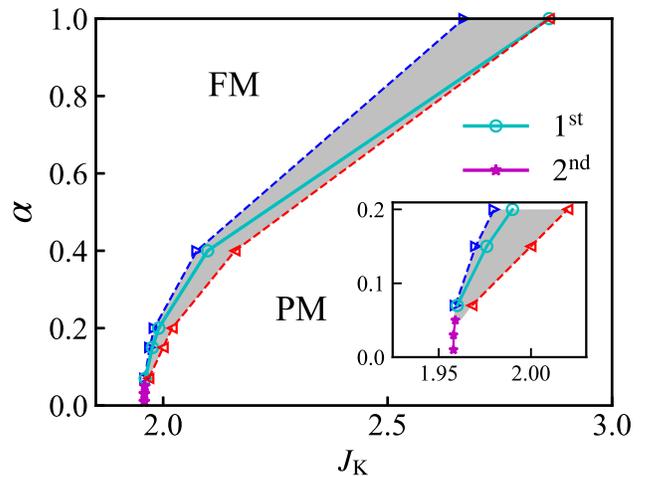}
	\end{center}
	\caption{The overall phase diagram on the $\alpha$-$J_{\rm K}$ plane for the anisotropic FM Kondo lattice model. The purple solid line represents the continuous (second-order)  QPT for large anisotropy. The shaded area marks the hysteresis region for the first-order QPT. The blue and red dashed lines denote its boundaries and the cyan solid line marks the energy crossing point of two coexisting phases. The inset shows an enlarged view for $\alpha\le0.2$.} 
	\label{fig5}
\end{figure}

Having established the order of the FM QPTs in the isotropic and quasi-1D cases, we now give more details on its evolution between the two limits. Figure~\ref{fig4} shows the calculated $m^2_\mathrm{FM}$ as a function of $J_\mathrm{K}$ for several intermediate values of $\alpha$. With decreasing $\alpha$, the hysteresis region of the first-order QPT is gradually reduced for $1\ge\alpha\ge 0.15$ and eventually becomes indiscernible within our numerical error for $\alpha<0.05$. Figure~\ref{fig5} summarizes the overall phase diagram of the ground state on the $\alpha$-$J_\mathrm{K}$ plane. The FM transition is continuous for sufficiently small $\alpha$ in the quasi-1D limit and first-order for $\alpha$ beyond a critical value $\alpha_c\simeq 0.05$. The value of $J_\mathrm{K}$ at the transition decreases for smaller $\alpha$. This is somewhat expected since a smaller $\alpha$ means a weaker  ferromagnetic coupling between the Kondo chains. Hence, a smaller Kondo coupling is enough to induce sufficient quantum fluctuations to destroy the FM alignment. It should be noted that since the width of the hysteresis region of the first-order transition is expected to be reduced in the limit of infinite bond dimension $D$ in our iPEPS calculations, the phase diagram presented in Fig.~\ref{fig5} should only be considered as correct in a qualitative sense. The results may be improved in the future by using more accurate full update \cite{Jordan2008Classical, Corboz2010Simulation, Phien2015Infinite} or variation update \cite{Vanderstraeten2016Gradient, Corboz2016Variational, Liao2019Differentiable}, or  increasing the virtual bond dimension with the help of symmetries \cite{Singh2010Tensor, Bauer2011Implementing, Singh2011Tensor, WEICHSELBAUM2012Non}. Nevertheless, our conclusion is consistent with previous results based on the large-$N$ Schwinger boson approximation \cite{wang2021unified} and provides additional support for the  anisotropy to be a possible general mechanism in causing FM quantum criticality in Kondo lattice systems.

So far, we have assumed the same anisotropy $\alpha_t=\alpha_\mathrm{H}=\alpha$ for the hopping parameter and the magnetic interaction. It has been argued previously that the Luttinger liquid property might cause the continuous FM QPT in the 1D limit because the soft modes driving the first-order transition are no longer present \cite{Brando2016Metallic}. On the other hand, the large-$N$ Schwinger boson calculations predicted a FM QCP with anisotropic magnetic interaction ($\alpha_H\ll 1$) and isotropic electron Fermi surfaces ($\alpha_t=1$) \cite{wang2021unified}. It is therefore interesting to clarify the role of the dimensionality of the electron degree of freedom. So we have also performed calculations by tuning $\alpha_t$ and $\alpha_\mathrm{H}$ separately. For fixed $\alpha_\mathrm{H}=1$, we find that reducing $\alpha_t$ from 1.0 to $0.01$ has no qualitative influence on the first-order FM QPT (not shown). Hence, the 1D property of the electron degree of freedom alone cannot suppress the first-order transition, and the magnetic anisotropy is crucial for realizing the FM QCP. If we fix $\alpha_t=1$ and change $\alpha_\mathrm{H}$ from 1.0 to a small value (e.g., 0.01), AFM correlations are found to gradually develop along the $x$-axis because of the induced AFM RKKY interaction which eventually overcomes the small FM $J_\mathrm{H}^x=\alpha_\mathrm{H} J_\mathrm{H}$. The FM long-range order is then replaced by a stripe AFM order with FM spin alignment along the $y$-axis and AFM alignment along the $x$-axis. We have firstly an FM-AFM transition before the AFM order is suppressed continuously to a PM upon further increasing $J_\mathrm{K}$. One may also tune $J_\mathrm{H}$ while keeping $\alpha_t=\alpha_\mathrm{H}=1.0$. Then the induced AFM RKKY interaction can become dominant along both axes for small $J_\mathrm{H}$, so we have first a N\'eel-type AFM order. In both cases, we do not have a direct FM-PM transition.

In conclusion, we have developed an iPEPS algorithm for the anisotropic Kondo lattice model and investigated the possibility of FM quantum criticality in two dimensions. Our calculations confirm the existence of a continuous FM QPT in quasi-1D systems with large hopping and magnetic anisotropy. It should be noted that spin anisotropy, crystal field effect, and multiple conduction bands \cite{WANG20211389Localized} may also play a role in realistic materials, which is beyond our model calculations and require more specific material study. Nevertheless, our work  highlights the importance of magnetic anisotropy for understanding FM quantum criticality and provides numerical confirmation of a possible generic mechanism for its presence in Kondo lattice systems. More elaborations along this line are needed to establish a microscopic theory of the FM quantum criticality. Our methods may be further extended to study other exotic physics such as spin-triplet superconductivity or combined effects of lattice geometry, spin-orbit coupling \cite{Kirkpatrick2020Ferromagnetic, miserev2022instability}, spin anisotropy \cite{Shen2020Strange}, and even disorder \cite{Sang2014Disorder} on the 2D Kondo lattice.

This work was supported by the National Key Research and Development Program of China (Grant No. 2017YFA0303103), the National Natural Science Foundation of China (NSFC Grants No. 12174429, and No. 11974397), and the Strategic Priority Research Program of the Chinese Academy of Sciences (Grant No. XDB33010100).

\end{document}


\title{Continuous ferromagnetic quantum phase transition on an anisotropic Kondo lattice\\
\vspace{0.2cm}
- Supplemental Material -}
\author{Jialin Chen}
\affiliation{Beijing National Laboratory for Condensed Matter Physics and Institute of
Physics, Chinese Academy of Sciences, Beijing 100190, China}
\affiliation{University of Chinese Academy of Sciences, Beijing 100049, China}
\author{Jiangfan Wang}
\affiliation{Beijing National Laboratory for Condensed Matter Physics and Institute of
Physics, Chinese Academy of Sciences, Beijing 100190, China}
\author{Danqing Hu}
\affiliation{Beijing National Laboratory for Condensed Matter Physics and Institute of
	Physics, Chinese Academy of Sciences, Beijing 100190, China}
\author{Yi-feng Yang}
\email[]{yifeng@iphy.ac.cn}
\affiliation{Beijing National Laboratory for Condensed Matter Physics and Institute of
Physics, Chinese Academy of Sciences, Beijing 100190, China}
\affiliation{University of Chinese Academy of Sciences, Beijing 100049, China}
\affiliation{Songshan Lake Materials Laboratory, Dongguan, Guangdong 523808, China}
\date{\today}

\maketitle

\renewcommand{\theequation}{S\arabic{equation}}
\renewcommand{\thefigure}{S\arabic{figure}}
\renewcommand{\bibnumfmt}[1]{[S#1]}

\begin{figure}[h]
	\begin{center}
		\includegraphics[width=8.2cm]{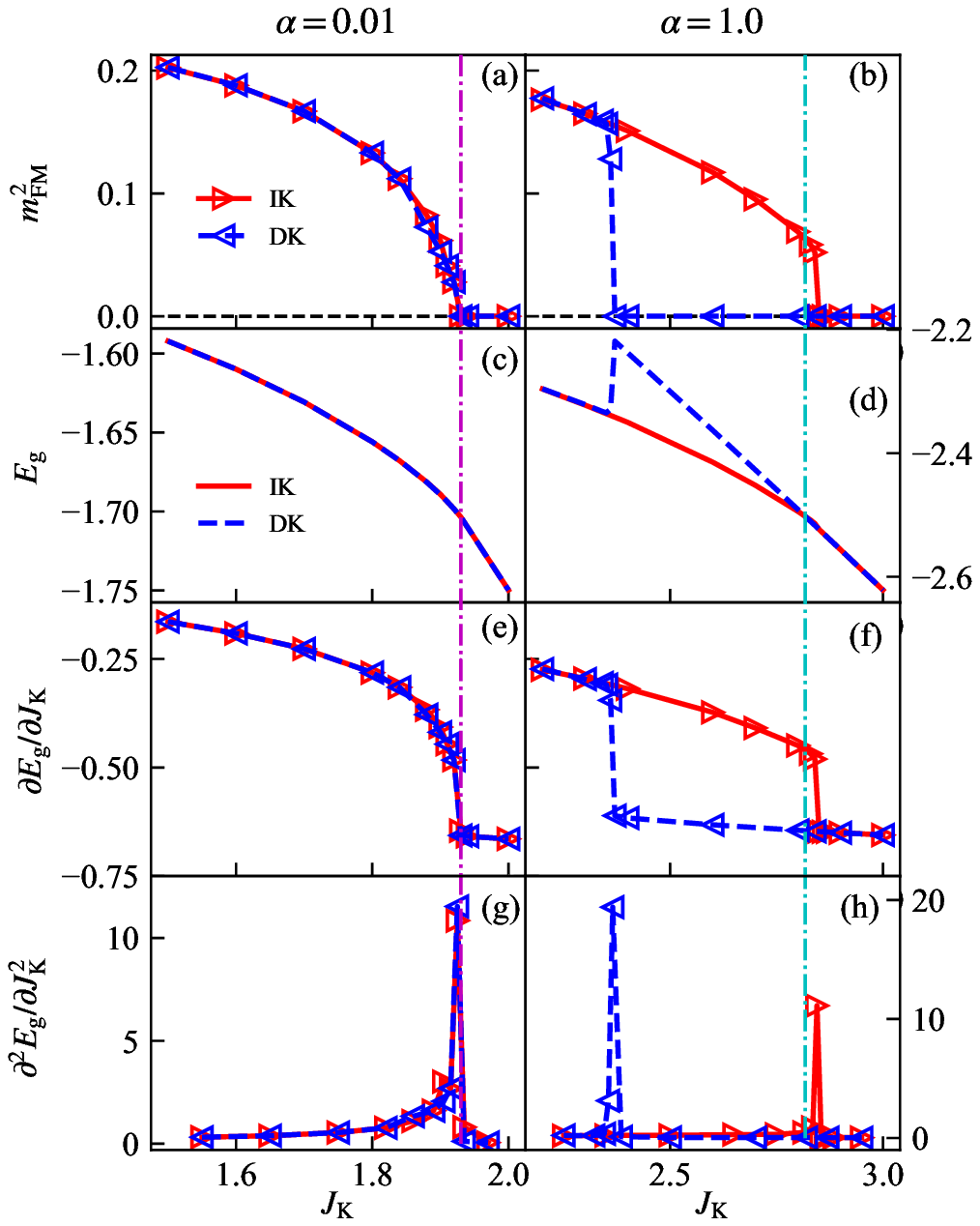}
	\end{center}
	\caption{Fast full update results for the isotropic ($\alpha=1.0$) and quasi-1D ($\alpha=0.01$) limits as a function of $J_\mathrm{K}$. The model parameters are the same as in Fig. 2 of the main text. The bond dimensions are $D=8$ and $\chi=64$.}
	\label{figS1}
\end{figure}

Here we provide iPEPS results with fast full update \cite{Phien2015Infinite}, which is a more accurate update algorithm than simple update but requires more computation time. Our calculations are hence limited to the bond dimensions $D=8$ and $\chi=64$. Figure \ref{figS1} shows the results for the isotropic ($\alpha=1.0$) and quasi-1D ($\alpha=0.01$) limits as in Fig. 2 of the main text with the same model parameters. The results are qualitatively the same, supporting that the FM transition is first order for $\alpha=1.0$ and second order for $\alpha=0.01$.

